\documentclass[aps,prl,twocolumn,groupedaddress,showpacs]{revtex4}

\usepackage{graphicx}% Include figure files
\usepackage{dcolumn}% Align table columns on decimal point
\usepackage{bm}% bold math
\usepackage{color}

\begin{document}

%%%%%%%%%%%%%%%%%%%%%%%%%%%%%% Title %%%%%%%%%%%%%%%%%%%%%%%%%%%%%%%%%%%
\title{Electric Dipolar Kondo Effect Emerging from Vibrating Magnetic Ion}

%%%%%%%%%%%%%%%%%%%%%%%%%%%% Author(s) %%%%%%%%%%%%%%%%%%%%%%%%%%%%%%%%%
\author{Takashi Hotta$^{1}$ and Kazuo Ueda$^{2}$}

%%%%%%%%%%%%%%%%%%%%%%%%% Organization(s) %%%%%%%%%%%%%%%%%%%%%%%%%%%%%%
\affiliation{$^1$Department of Physics, Tokyo Metropolitan University,
Hachioji, Tokyo 192-0397, Japan \\
%$^2$Advanced Science Research Center,
%Japan Atomic Energy Agency, Tokai, Ibaraki 319-1195, Japan \\
$^2$Institute for Solid State Physics,
University of Tokyo, Kashiwa, Chiba 277-8581, Japan}

\date{\today}

%%%%%%%%%%%%%%%%%%%%%%%%%%%%% Abstract %%%%%%%%%%%%%%%%%%%%%%%%%%%%%%%%%
\begin{abstract}
When a magnetic ion vibrates in a metal, it inevitably introduces
a new channel of hybridization with conduction electrons
and in general, the vibrating ion induces electric dipole moment.
In such a situation,
we find that magnetic and non-magnetic Kondo effects alternatively
occur due to the screening of spin moment and electric dipole moment
of vibrating ion.
In particular, electric dipolar two-channel Kondo effect is found
to occur for weak Coulomb interaction.
We also show that magnetically robust heavy-electron state appears
near the fixed point of electric dipolar two-channel Kondo effect.
We believe that the {\it vibrating} magnetic ion
opens a new door in the Kondo physics.
\end{abstract}

%%%%%%%%%%%%%%%%%%%%%%%%%%%%%%% Pacs %%%%%%%%%%%%%%%%%%%%%%%%%%%%%%%%%%%

% 75.20.Hr Local moment in compounds and alloys; Kondo effect, valence
%          fluctuations, heavy fermions
% 75.40.Cx Static properties (order parameter, static susceptibility, heat
%          capacities, critical exponents, etc.)
% 71.27.+a Strongly correlated electron systems; heavy fermions

\maketitle

%%%%%%%%%% introduction %%%%%%%%%%

It has been widely recognized that Kondo phenomena generally appear
when localized entity with internal degrees of freedom
is coupled with conduction electrons.
Concerning the original problem of resistance minimum phenomenon
in metals with magnetic impurities,
Kondo has actually shown it by quantum-mechanical calculations
for scattering amplitude of electrons due to magnetic impurities
\cite{Kondo}.
Then, it has been revealed that the singlet state is formed
from local magnetic moment
due to the coupling with conduction electrons \cite{Yosida}.
After the understanding of the Kondo effect in the dilute magnetic
impurity system, interests of researchers have moved to
the impurity with complex degrees of freedom.

One research direction has been found in the explicit consideration
of orbital degree of freedom of localized electron.
Coqblin and Schrieffer have derived exchange interactions
from the multiorbital Anderson model \cite{Coqblin}.
Then, the concept of multi-channel Kondo effect has been developed
on the basis of such exchange interactions \cite{Nozieres},
as a potential source of non-Fermi liquid phenomena.
Such non-Fermi liquid properties have been pointed out
also in a two-impurity Kondo system \cite{Jones1,Jones2}.
Concerning the reality of two-channel Kondo effect,
Cox has pointed out the existence of two screening channels
in the case of quadrupole degree of freedom in a cubic uranium compound
with non-Kramers doublet ground state \cite{Cox}.

Another possibility is non-magnetic Kondo effect with phonon origin.
First Kondo has considered Kondo-like behavior in a two-level
system \cite{Kondo2}.
The two-level Kondo system has been proven to exhibit
the same behavior as the magnetic Kondo effect \cite{Vladar}.
Yu and Anderson have discussed the phononic Kondo effect
from a different viewpoint \cite{Yu-Anderson}.
They have pointed out that the scattering process between spinless
$s$-wave conduction electron to $p$-wave one is produced by
ion displacement.

Recently, the Kondo effect with phonon origin has attracted
renewed attention due to active experimental investigations
on cage structure materials, in which a guest ion is vibrating
in a cage composed of relatively light atoms.
We believe that the vibration of magnetic ion in the cage
provides a new ingredient in the Kondo physics,
since dynamical aspects of magnetic impurity have been
considered unsatisfactorily in the Kondo problem.
In fact, quite recently, the two-channel Kondo effect has been
confirmed in the model for vibrating magnetic impurity
\cite{Dagotto,Yashiki1,Yashiki2,Yashiki3}.
Also for the promotion of our understanding on magnetically robust
heavy-electron phenomenon observed in cage compound \cite{Sanada},
we further develop the Kondo physics of vibrating magnetic impurity.

In this Letter, we analyze a two-channel conduction electron system
hybridized with vibrating magnetic ion
by using a numerical renormalization group technique.
We confirm magnetic and non-magnetic
Kondo effects originating from the screening of spin
and electric dipolar moments, respectively,
by evaluating entropy and susceptibilities
for spin and electric dipole moments.
Near the fixed point for electric dipolar
two-channel Kondo effect, we find magnetically robust heavy-electron
state from the direct evaluation of the Sommerfeld constant.

%%%%%%%%%% model and technique %%%%%%%%%%

Let us consider a two-channel conduction electron system
hybridized with vibrating magnetic impurity
\cite{Yashiki1, Yashiki2, Yashiki3}.
In the unit of $\hbar$=$k_{\rm B}$=$1$,
the Hamiltonian is given by
\begin{eqnarray}
  H &=& \sum_{\bm{k},\sigma}\left[
  \varepsilon_{\bm{k}}(c_{\bm{k}s\sigma}^{\dag} c_{\bm{k}s\sigma}
  +c_{\bm{k}p\sigma}^{\dag} c_{\bm{k}p\sigma})\right] \nonumber \\
  &+& \sum_{\bm{k},\sigma}\left[
   V_0(c_{\bm{k}s\sigma}^{\dag}f_{\sigma}+{\rm h.c.})
   +gx(c_{\bm{k}p\sigma}^{\dag}f_{\sigma}+{\rm h.c.})\right] \\
  &+& U n_{\uparrow}n_{\downarrow}+E_f(n_{\uparrow}+n_{\downarrow})
+\omega x^2/2+p^2/2,\nonumber
\end{eqnarray}
where $\varepsilon_{\bm{k}}$ denotes conduction electron dispersion,
$c_{\bm{k}\ell\sigma}$ indicates the annihilation operator for
conduction electron with momentum $\bm{k}$, angular momentum $\ell$,
and spin $\sigma$,
$f_{\sigma}$ is the annihilation operator for localized electron
with spin $\sigma$,
$n_{\sigma}$=$f^{\dag}_{\sigma}f_{\sigma}$,
$U$ is the Coulomb interaction between localized electrons,
$E_f$ denotes the local $f$-level energy,
$V_0$ is the hybridization between $s$-channel conduction and
localized $f$ electrons,
$g$ is the electron-vibration coupling,
$x$ denotes the ion displacement,
$p$ indicates the corresponding canonical momentum,
and $\omega$ is the vibration frequency.
Note that we set the reduced mass of vibration as unity.
In order to consider the symmetric case,
we set $E_f$=$-U/2$ throughout this paper.

In this paper, we analyze the model with the use of
a numerical renormalization group (NRG) method \cite{NRG}.
We introduce a cut-off $\Lambda$
for the logarithmic discretization of the conduction band.
Due to the limitation of computer resources,
we keep $N$ low-energy states.
In this paper, we set $\Lambda$=5 and $N$=$5000$$\sim$$10000$.
Note that the temperature $T$ is defined as
$T$=$\Lambda^{-(i-1)/2}$ in the NRG calculation,
where $i$ is the number of the renormalization step.
The phonon basis is truncated at a finite number $N_{\rm ph}$,
which is set as $N_{\rm ph}$=300 in this paper.

It is convenient to introduce phonon operators $a$ and $a^{\dag}$
through the relation of $x$=$(a+a^{\dag})/\sqrt{2\omega}$.
Then, we define the magnitude of phonon-assisted hybridization
as $V_1$=$g/\sqrt{2\omega}$.
The energy unit is the half of the conduction electron band $D$,
which is set as unity.
As for parameters, we set $V_0$=$\omega$=$0.2$
and we change the values of $U$ and $V_1$.

%%%%%%%%%%%%%%%%%%%%%%%% Fig.1 %%%%%%%%%%%%%%%%%%%%%%%%%%%%%%%%%%%%%
\begin{figure}[t]
\includegraphics[width=5cm]{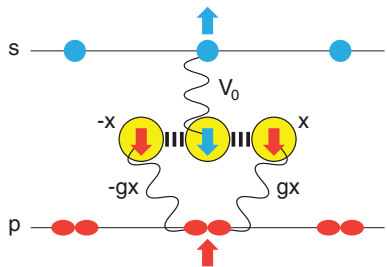}
\caption{(Color online) Schematic view of the situation described
by the model Hamiltonian (1).
}
\end{figure}
%%%%%%%%%%%%%%%%%%%%%%%%%%%%%%%%%%%%%%%%%%%%%%%%%%%%%%%%%%%%%%%%%%%%%

%%%%%%%%%% Schematic view %%%%%%%%%%

First let us visualize the situation described by the present model.
In Fig.~1, the two horizontal lines symbolically denote $s$
and $p$ channels which are hybridized with magnetic impurity.
As shown in $H$, $s$-channel electrons are hybridized with
localized electron in a standard manner,
while the hybridization process between
$p$-channel and localized electrons is assisted by phonons.
Note that parities of $s$- and $p$-channel electrons are different,
as easily understood from the values of angular momenta.
Namely, $s$- and $p$-channel electrons possess even and odd parities,
respectively.
Since the parity for ion displacement $x$ is odd,
it is natural that the phonon-assisted hybridization
occurs only for $p$ channel.

When magnetic ion stops at an origin, we consider only
the screening of impurity spin moment by $s$-channel electrons,
leading to the conventional Kondo effect.
However, when the ion is vibrating as shown in Fig.~1,
there occurs another screening process due to $p$-channel electrons.
In such a situation, in addition to the screening of spin moment,
the electric dipole moment in proportion to the displacement $x$
should be also screened by conduction electrons,
leading to non-magnetic Kondo effect.
Intuitively, we understand that
the main screening channel is converted between $s$ and $p$
due to the balance between $V_0$ and $V_1$.
In fact, the phases with different quantum numbers in relation
with parity have been found to be converted between the regions of
large and small $V_1$.
Then, two-channel Kondo effect has been confirmed to occur
just at the boundary between those two phases
\cite{Dagotto,Yashiki1,Yashiki2}.
However, another conversion between magnetic and non-magnetic Kondo
effects controlled by the balance between $U$ and $V_1$ has not been
clarified yet.
Then, in this paper, we unveil such a new point in relation with
magnetically robust heavy electron state.

%%%%%%%%%%%%%%%%%%%%%%%% Fig.2 %%%%%%%%%%%%%%%%%%%%%%%%%%%%%%%%%%%%%
\begin{figure}[t]
\includegraphics[width=8.5cm]{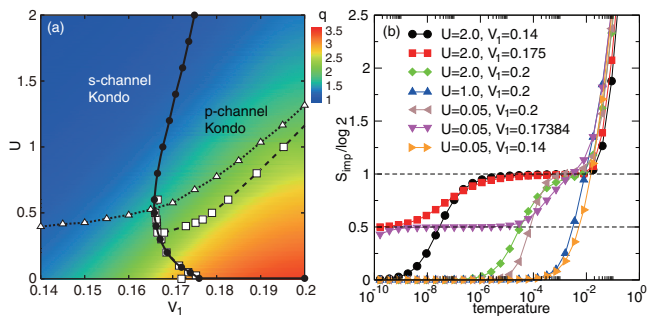}
\caption{(Color online)
(a) Phase diagram on the $V_1$-$U$ plane for $V_0$=$\omega$=$0.2$.
The color gradation indicates the magnitude of
average displacement $q$.
Solid curve with solid circles denote the two-channel Kondo line.
Dotted curve with open triangles and broken curve with open squares
denote $d^2 q/d U^2$=$0$ and $d^2 q/d V_1^2$=$0$, respectively.
As for the definition of $q$, see the main text.
(b) Entropy vs. temperature.
}
\end{figure}
%%%%%%%%%%%%%%%%%%%%%%%%%%%%%%%%%%%%%%%%%%%%%%%%%%%%%%%%%%%%%%%%%%%%%

%%%%%%%%%% results %%%%%%%%%%

Now we explain the phase diagram in Fig.~2(a).
As mentioned above, the solid curve indicates the boundary
between two phases with different quantum numbers,
which are obtained by the $s$- and $p$-channel
Kondo screening, respectively.
On the boundary curve between two phases,
the two-channel Kondo effect is realized.
Since no phase conversion occurs on the line of $U$=$0$,
the boundary curve asymptotically approaches the line of $U$=$0$.
Note that low-energy spectra of three fixed points of
$s$-channel, $p$-channel, and two-channel Kondo states
have been revealed in Refs.~\cite{Yashiki1} and \cite{Yashiki2}.
New results of this paper are color gradation and
other broken and dotted curves.
Their meanings will be discussed later.

Next we discuss the NRG results of entropy.
In Fig.~2(b), we show the change of entropy along the lines
of $U$=$2$, $V_1$=$0.2$, and $U$=$0.05$.
On the line of $U$=$2$, for $V_1$=$0.14$ and $0.2$,
we find the plateaus of $\log 2$, which are eventually released
at low enough temperatures.
At $(V_1, U)$=$(0.175, 2)$, we observe the entropy of $0.5\log 2$
at low temperatures, which is the signal of two-channel Kondo effect.
When we decrease the value of $U$ from $U$=$2$
on the line of $V_1$=$0.2$,
we observe the increase of the Kondo temperature $T_{\rm K}$,
which is characterized by the release of entropy $\log 2$.
This is quite natural from the viewpoint of the conventional
magnetic Kondo effect.
Thus, we deduce that the Kondo effect on the line of $U$=$2$
as well as in the region of $U$$\agt$$1$ on the line of $V_1$=$0.2$
is originating from the screening of impurity spin moment.
However, for $U$$\alt$$1$, a plateau of $\log 2$ again appears.
The temperature region of $\log 2$ is wider for smaller $U$
and $T_{\rm K}$ is decreased with the decrease of $U$.
Such behavior is contradictory to the conventional
magnetic Kondo effect.
Furthermore, at $(V_1, U)$=$(0.17384, 0.05)$,
we again observe a clear plateau of entropy of $0.5\log 2$,
but it is difficult to understand the origin of this two-channel
Kondo behavior only from the result of entropy.

In order to clarify what quantity is screened,
we evaluate susceptibilities for magnetic and
electric dipole moments, which are, respectively,
defined by
\begin{eqnarray}
  \chi_M=\int_0^{1/T} d\tau \langle M(\tau) M \rangle,~~
  \chi_P=\int_0^{1/T} d\tau \langle P(\tau) P \rangle.
\end{eqnarray}
Here $\langle \cdots \rangle$ denotes the operation
to take thermal average,
$M(\tau)$=$e^{H\tau}Me^{-H\tau}$,
$M$=$g_s\mu_{\rm B}(n_{\uparrow}-n_{\downarrow})/2$,
and $P$=$Ze(a+a^{\dag})/\sqrt{2\omega}$,
where $g_s$ is the electron $g$-factor which is set as $g_s$=$2$,
$\mu_{\rm B}$ is the Bohr magneton,
$Z$ is valence number of ion,
and $e$ is electric charge.
In the actual calculations, we normalize them as
$\chi_M/\mu_{\rm B}^2$ and $\chi_P/(Z^2e^2/2\omega)$.

In Fig.~3(a), we show $T\chi_M$ for the same values of
$U$ and $V_1$ in Fig.~2(b).
As we have deduced above, on the line of $U$=$2$,
we observe the decrease of $T\chi_M$ around $T_{\rm K}$.
Note that the curves of $T\chi_M$ for $V_1$=$0.14$
and $V_1$=$0.2$ at $U$=$2$ are quite similar at low temperatures,
when $T$ is rescaled by $T_K$.
On the other hand, the curve of $T\chi_M$ for $(V_1, U)$=$(0.175, 2)$
is apparently different from those for $V_1$=$0.14$ and $V_1$=$0.2$.
It is due to the non-Fermi liquid behavior in the two-channel Kondo
effect, leading to $\chi_M$$\sim$$-\log T$.
%In the present calculation.
This logarithmic correction in $\chi_M$
is clearly observed.
Along the lines of $V_1$=$0.2$ and $U$=$0.05$,
$T\chi_M$ decreases at relatively high temperature,
but the release of the entropy does not seem to correspond
to the temperature at which $T\chi_M$ is decreased.

Now we turn our attention to the results for
$T\chi_P$ in Fig.~3(b).
The entropy release for $(V_1, U)$=$(0.2, 0.05)$
corresponds to the decrease of $T\chi_P$,
indicating the occurrence of the Kondo effect concerning
electric dipole moment.
Note that $\chi_P$ is related to phonon Green's function.
In the strong electron-phonon coupling region,
the center of oscillation is
shifted either right or left, leading to $\chi_P$$\propto$$q^2/T$,
where $q$=$\sqrt{\langle (a+a^{\dag})^2 \rangle}$.
Thus, for small $U$ and large $V_1$ region,
we expect that $T\chi_P$ becomes constant at high temperatures,
as found in Fig.~3(b).

When $T$ is decreased, $T\chi_P$ is decreased from
the constant value due to the Kondo screening of electric dipole
moment and it eventually goes to zero at $T$=0.
This can be called the parity Kondo effect, since
near degeneracy with different phonon parities
characterizes the electric dipole,
which is coupled with conduction electron parity,
leading to the non-degenerate ground state with fixed total parity.
Note that the total parity is specified by 0 or 1,
depending on $U$ and $V_1$.
At low enough temperatures, we find
$T\chi_P$=$2T/{\tilde \omega}_1$,
where ${\tilde \omega}_1$ is renormalized phonon energy
smaller than $\omega$.
From the local Fermi-liquid theory, we find
$T_{\rm K} \propto {\tilde \omega}_1$,
but the proportional coefficient is suppressed
by the polaron effect in comparison with
the conventional magnetic Kondo effect.
Then, $T\chi_P$ decreases rapidly around $T_{\rm K}$
in sharp contrast to $T\chi_M$ around $T_{\rm K}$.
The shape of $T\chi_P$ at $(V_1, U)$=$(0.2, 1.0)$ around $T_{\rm K}$
is quite similar to that for $(0.2, 0.05)$ \cite{comment},
when $T$ is rescaled by $T_{\rm K}$.
For the case of $(0.2, 1.0)$,
corresponding to the competing region of
magnetic and electric dipolar Kondo effects,
we do not find any significant structure in the entropy and $T\chi_M$,
but in the decrease of $T\chi_P$ from the shoulder,
the enhanced signal can be observed
due to the polaron effect in the phonon matrix element of $\chi_P$.

%%%%%%%%%%%%%%%%%%%%%%%% Fig.3 %%%%%%%%%%%%%%%%%%%%%%%%%%%%%%%%%%%%%
\begin{figure}[t]
\includegraphics[width=8.5cm]{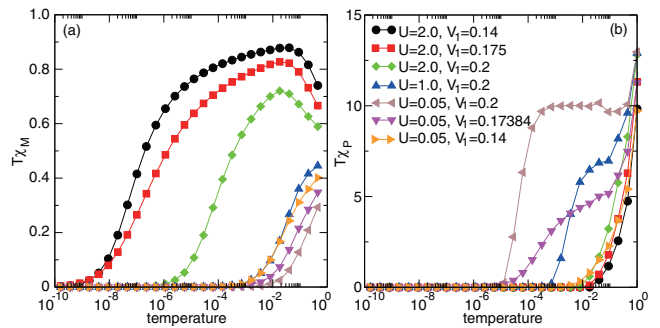}
\caption{(Color online)
Susceptibilities for (a) magnetic and (b) electric dipole
moments as functions of temperature.
}
\end{figure}
%%%%%%%%%%%%%%%%%%%%%%%%%%%%%%%%%%%%%%%%%%%%%%%%%%%%%%%%%%%%%%%%%%%%%

We remark that the two-channel Kondo effect also
occurs due to the screening of electric dipole moment \cite{note}.
In fact, at $(V_1, U)$=$(0.17384, 0.05)$,
we find a plateau of entropy $0.5 \log 2$
and the significant decrease of
$T\chi_P$ around the corresponding temperature.
Note that the shape of $T\chi_P$ at $(0.17384, 0.05)$ is
different from those for $(0.2, 0.05)$ and $(0.2, 1.0)$.
We observe the smooth change from $T\chi_P$=constant to
$\chi_P$=constant in this case.
Since the electric dipole moment is not perfectly screened,
vibration still remains and thus,
we intuitively obtain $\chi_P$=$2/{\tilde \omega}_2$
in the two-channel electric dipolar Kondo regime,
where ${\tilde \omega}_2$ is another renormalized phonon energy,
which is different from ${\tilde \omega}_1$.
It is one of future tasks to explain the difference
between ${\tilde \omega}_1$ and ${\tilde \omega}_2$
by overcoming the difficulty to estimate ${\tilde \omega}_1$
with high precision in the NRG calculation.

Here let us go back to Fig.~1(a).
We have found that the electric dipole moment is screened
in the region of large $V_1$ and small $U$.
In order to visualize the change of the screened moment,
we depict $q$ as the color gradation in Fig.~1(a).
For large $V_1$ and small $U$, we find the red region
with large $q$,
in which electric dipolar Kondo effect occurs.
On the other hand, there occurs magnetic Kondo effect
in the blue region of small $V_1$ and large $U$.
As a guide of the boundary between magnetic and non-magnetic
Kondo regions,
we plot inflection points of $d^2 q/d U^2$=$0$ (dotted)
and $d^2 q/d V_1^2$=$0$ (broken) in Fig.~1(a).
For large $V_1$, two curves run in the green area
between blue and red regions.
Another broken curve appears on the two-channel Kondo line
for $U$$\alt$$0.6$, except for $U$=$0$,
suggesting the electric dipolar two-channel Kondo region.
Note, however, that the inflection points seem to lose
the meaning of boundary in the $s$-channel Kondo region,
since renormalized Fermi chain is realized
in the region of small $U$ and small $V_1$
\cite{Yashiki1, Yashiki2}.
In the small area of yellow and orange
of the $s$-channel Kondo region,
there also occurs non-magnetic Kondo effect,
which is interpreted as
the Yu-Anderson Kondo effect \cite{Yu-Anderson}.

Now let us discuss the magnetically robust heavy electron state
by the Sommerfeld constant $\gamma$.
For the purpose, we add the Zeeman term
$H_{\rm Z}$=$g_s \mu_{\rm B} H (n_{\uparrow}-n_{\downarrow})/2$,
where $H$ is an applied magnetic field, to the model (1).
Then, we evaluate $\gamma$ at $T$=$\Lambda^{-14}$,
the lowest temperature at which we can arrive
in the present NRG calculations.
In Fig.~4(a), we show the results for $\gamma$
in the unit of ${\rm mJ} / {\rm mol} \cdot {\rm K}^2$
with $D$=$1$ eV.
We find that $\gamma$ shows divergent behavior around
the two-channel Kondo fixed points,
shown by the solid curve in Fig.~1(a).
On the line of $U$=$0$, we do not find the phase conversion,
but for $V_1$$\agt$$0.17$, the two-channel Kondo line
exists in the extreme vicinity of $U$=$0$.
Thus, $\gamma$ becomes very large for $U$=$0$ and $V_1$$\agt$$0.17$.
In the vicinity of the two-channel Kondo line,
we find the enhancement of $\gamma$ due to the non-Fermi liquid
properties, irrespective of $U$.
For the parameters away from the two-channel Kondo fixed points
for $U$$>$$0.05$, $\gamma$ in the region of the magnetic Kondo effect
is relatively large in comparison with that of
the non-magnetic Kondo effect of electric dipole moment.

%%%%%%%%%%%%%%%%%%%%%%%% Fig.4 %%%%%%%%%%%%%%%%%%%%%%%%%%%%%%%%%%%%%
\begin{figure}[t]
\includegraphics[width=8.5cm]{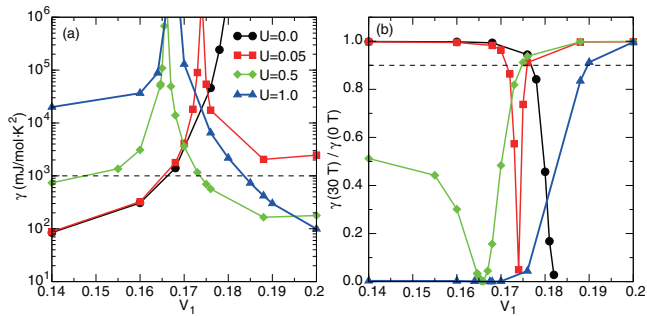}
\caption{(Color online)
(a) Sommerfeld constant vs. $V_1$ for $H$=$0$ and (b) the ratio
of the Sommerfeld constants vs. $V_1$.
}
\end{figure}
%%%%%%%%%%%%%%%%%%%%%%%%%%%%%%%%%%%%%%%%%%%%%%%%%%%%%%%%%%%%%%%%%%%%%

In Fig.~4(b), we show the ratio of $\gamma$'s at $H$=0 T and 30 T.
If this ratio is near the unity, we judge that $\gamma$ is
magnetically robust.
As we easily imagine, in the Kondo effect concerning the
electric dipole moment, $\gamma$ does not depend sensitively
on the magnetic field.
As for criteria of the magnetically robust heavy electron state,
we consider the conditions of $\gamma$$>$$1000$ and
$\gamma (30 {\rm T})/\gamma (0 {\rm T})$$>$$0.9$,
shown by horizontal broken lines in Figs.~4(a) and (b).
After the NRG calculations, we find that
the magnetically robust heavy electron state appears
in the region of $V_1$$\agt$$0.17$ and $U$$\alt$$0.1$,
except for the narrow region near the two-channel Kondo line.
Note that the ratio is strongly suppressed due to the divergent
behavior of $\gamma (0 {\rm T})$.
That region is included in the non-magnetic Kondo effect
region with orange and red in Fig.~1(a).
In comparison with the mechanism of magnetically robust
large $\gamma$ on the basis of
the charge Kondo effect \cite{Hotta},
it seems to be easier to obtain large $\gamma$
in the present scenario based on the non-magnetic
two-channel Kondo effect.

%%%%% Summary %%%%%

In summary, we have analyzed the two-channel conduction
electron model with vibrating magnetic ion.
We have found two types of Kondo effects due to the
alternative screening of magnetic and electric dipole moments.
Near but not exactly on the two-channel Kondo line
with electric dipolar origin,
we have found that $\gamma$ is magnetically robust.
Non-magnetic electric dipolar Kondo behavior is expected
to be observed in cage-structure materials.
In particular, magnetically robust non-Fermi liquid behavior
is an interesting possibility.

%%%%% Acknowledgement %%%%%

This work was supported by KAKENHI (20102008).
The computation in this work has been done using the facilities
of the Supercomputer Center of Institute for Solid State Physics,
University of Tokyo.

%%%%%%%%%%%%%%%%%%%%%%% references %%%%%%%%%%%%%%%%%%%%%%%%%%%%%%%%%


\begin{thebibliography}{99}

\bibitem{Kondo}
J. Kondo, Prog. Theor. Phys. {\bf 32}, 37 (1964).

\bibitem{Yosida}
K. Yosida, Phys. Rev. {\bf 147}, 223 (1966).

\bibitem{Coqblin}
B. Coqblin and J. R. Schrieffer,
Phys. Rev. {\bf 185}, 847 (1969).

\bibitem{Nozieres}
Ph. Nozi\'eres and A. Blandin,
J. Physique {\bf 41}, 193 (1980).

\bibitem{Jones1}
B. A. Jones and C. M. Varma,
Phys. Rev. Lett. {\bf 58}, 843 (1987).

\bibitem{Jones2}
B. A. Jones {\it et al.},
%C. M. Varma, and J. W. Wilkins,
Phys. Rev. Lett. {\bf 61}, 125 (1988).

\bibitem{Cox}
D. L. Cox, Phys. Rev. Lett. {\bf 59}, 1240 (1987).

\bibitem{Kondo2}
J. Kondo, Physica B+C {\bf 84}, 40 (1976);
Physica B {\bf 84}, 207 (1976).

\bibitem{Vladar}
K. Vladar and A. Zawadowski,
Phys. Rev. B {\bf 28}, 1564 (1983);
{\it ibid} {\bf 28}, 1582 (1983).

\bibitem{Yu-Anderson}
C. C. Yu and P. W. Anderson,
Phys. Rev. B {\bf 29}, 6165 (1984).

\bibitem{Dagotto}
L. G. G. V. Dias da Silva and E. Dagotto,
Phys. Rev. B {\bf 79}, 155302 (2009).

\bibitem{Yashiki1}
S. Yashiki {\it et al.},
%S. Kirino, and K. Ueda,
J. Phys. Soc. Jpn. {\bf 79}, 093707 (2010).

\bibitem{Yashiki2}
S. Yashiki {\it et al.},
%S. Kirino, K. Hattori, and K. Ueda,
J. Phys. Soc. Jpn. {\bf 80}, 064701 (2011).

\bibitem{Yashiki3}
S. Yashiki and K. Ueda,
J. Phys. Soc. Jpn. {\bf 80}, 084717 (2011).

\bibitem{Sanada}
S. Sanada {\it et al.},
%Y. Aoki, H. Aoki, A. Tsuchiya, D. Kikuchi, H. Sugawara,
%and H. Sato,
J. Phys. Soc. Jpn. {\bf 74}, 246 (2005).

\bibitem{NRG}
K. G. Wilson, Rev. Mod. Phys. {\bf 47}, 773 (1975).
%H. R. Krishna-murthy, J. W. Wilkins, and K. G. Wilson,
%Phys. Rev. B {\bf 21}, 1003 (1980).

\bibitem{comment}
Since $T_{\rm K}$ is determined
by the tunneling matrix element
between the polaron doublet whose energy
is reduced by $U$, $T_{\rm K}$ is increased
with the increase of $U$.

\bibitem{note}
In actual cage materials,
we roughly estimate $V_1$ in the order of 0.01 eV,
which does not seem to be large enough to observe electric
dipolar Kondo effect.
However, in actuality, anharmonicity exists in the vibration.
When we consider the effect of anharmonicity \cite{Yashiki3},
the polaron binding energy is increased and
the electric dipolar Kondo effect is expected
to occur more easily.

\bibitem{Hotta}
T. Hotta, J. Phys. Soc. Jpn. {\bf 77}, 103711 (2008).

\end{thebibliography}
\end{document}